\documentclass[twocolumn,amsmath,amssymb]{revtex4}

\begin{document}

\title{Pseudospectral Algorithms for Solving\\
Nonlinear Schr\"odinger Equation in 3D}

\author{A. A. Skorupski}

\email[]{askor@fuw.edu.pl}

\affiliation{Department of Theoretical Physics, So{\l}tan Institute for Nuclear Studies\\
Ho\.za 69, 00--681, Warsaw, Poland}

\date{\today}

\begin{abstract}
Three pseudospectral algorithms are described (Euler, leapfrog and trapez) for solving
numerically the time dependent nonlinear Schr\"odinger equation in one, two or three
dimensions. Numerical stability regions in the parameter space are determined for the
cubic nonlinearity, which can be easily extended to other nonlinearities. For the first
two algorithms, maximal timesteps for stability are calculated in terms of the maximal
Fourier harmonics admitted by the spectral method used to calculate space derivatives.
The formulas are directly applicable if the discrete Fourier transform is used, i.e. for
periodic boundary conditions. These formulas were used in the relevant numerical programs
developed in our group.
\end{abstract}

\maketitle

\section{Introduction}

The nonlinear Schr\"odinger (NLS) equation  in one, two or three dimensions
is a commonly used model in various branches of physics, e.g. in plasma
theory \cite{infrow}, optics \cite{kerr} or condensed matter theory \cite{feynm}.
In these applications
the cubic form of the nonlinear term is used, and our analysis will be
pertinent to this form. However, the generalization to other nonlinearities
will usually be straighhtforward. To make the analysis directly applicable
to various situations we assume the following form of the NLS equation:
\begin{equation}\label{nls}
{i}\bar{u}_{\bar{t}} +
A \bar{u}_{\bar{x}\bar{x}} +
B \bar{u}_{\bar{y}\bar{y}} +
C \bar{u}_{\bar{z}\bar{z}} +
D |\bar{u}|^2 \bar{u} +
\bar{E} \bar{u} = 0 \, ,
\end{equation}
where $\bar{t}$ denotes time, and $\bar{x}, \, \bar{y}, \, \bar{z}$ are
space coordinates; subscripts of $\bar{u}$ denote partial derivatives.
The coefficients $A$ ($\neq 0$), $B$, $C$, $D$ ($\neq 0$) are assumed to
be real, but $\bar{E}$ can be complex, $\bar{E} = \bar{E}_{\text{r}} +
i \bar{E}_{\text{i}}$. The unknown function will be defined
in a box:
\begin{eqnarray}
&&-L_{\bar{x}} \le \bar{x} \le L_{\bar{x}} \, ,\quad
-L_{\bar{y}} \le \bar{y} \le L_{\bar{y}} \, ,\quad
-L_{\bar{z}} \le \bar{z} \le L_{\bar{z}} \, ,\nonumber\\
&&\label{box}
\end{eqnarray}
with periodic boundary conditions: $\bar{u}(-L_{\bar{x}}, \ldots) =
\bar{u}(L_{\bar{x}}, \ldots)$, etc. These conditions will be fulfilled
exactly for solutions periodic in $\bar{x}$, $\bar{y}$ and $\bar{z}$
but approximatelly also for solitary solutions, which are exponentially
small at the boundaries. Non-periodic boundary conditions require special treatment,
see the end of Sec.~III.

\section{Transformation to machine units}

For the machine variables, which will be used in programming, we will
drop the bars used in (\ref{nls}) to denote the original (physical)
variables. Both types of variables will be linearly related to each
other so as to make the NLS equation in machine units as simple as possible.
As the Fourier transform of a periodic function with period $P$ is the
simplest for $P = 2 \pi$, we first transform the space intervals in
(\ref{box}) into the intervals $[0, \, 2\pi]$ by putting
\begin{eqnarray}
x & = & \alpha_x (\bar{x} + L_{\bar{x}}), \qquad \alpha_x =
\frac{\pi}{L_{\bar{x}}} \, , \nonumber\\
y & = & \alpha_y (\bar{y} + L_{\bar{y}}), \qquad \alpha_y =
\frac{\pi}{L_{\bar{y}}} \, , \label{vtrans}\\
z & = & \alpha_z (\bar{z} + L_{\bar{z}}), \qquad \alpha_z =
\frac{\pi}{L_{\bar{z}}} \, , \nonumber
\end{eqnarray}
Furthermore we normalize the time and the unknown function
($\alpha_t, \, \alpha_u > 0$)
\begin{equation}\label{vnorm}
t  =  \alpha_t \bar{t} \, , \qquad
u  =  \alpha_u \bar{u} \, .
\end{equation}
With these transformations Eq. (\ref{nls}) becomes
\begin{eqnarray}
&&i u_t + \frac{A\alpha_x^2}{\alpha_t}
\Bigl[ u_{xx} + \frac{B}{A} \Bigl(\frac{L_{\bar x}}{L_{\bar y}}\Bigr)^2 u_{yy}
+ \frac{C}{A}\Bigl(\frac{L_{\bar x}}{L_{\bar z}}\Bigr)^2 u_{zz}\Bigr]\nonumber\\
&&+ \frac{D}{\alpha_u^2 \alpha_t} |u|^2 u
+ \frac{\bar E}{\alpha_t} u = 0 \, .\label{nlsmu}
\end{eqnarray}
We can choose $\alpha_t$ and $\alpha_u$ so that $|A|\alpha_x^2/\alpha_t
=1$ and $|D|/(\alpha_u^2 \alpha_t) = 1$, leading to
\begin{equation}\label{atau}
\alpha_t  =  |A|\alpha_x^2 \, ,\qquad
\alpha_u  =  \frac{\sqrt{|D/A|}}{\alpha_x} \, .
\end{equation}
With this choice Eq. (\ref{nlsmu}) can be written
\begin{equation}
u_t = F[u] \, ,\label{eveq}
\end{equation}
where
\begin{eqnarray}
F[u] &=& i \bigl[\text{sgn}(A) (u_{xx} + c_y u_{yy} + c_z u_{zz})\nonumber\\
&&+ \text{sgn}(D) |u|^2 u + E u \bigr],\label{rhs}
\end{eqnarray}
\begin{eqnarray}
c_y &=& \frac{B}{A}\Bigl( \frac{L_{\bar{x}}}{L_{\bar{y}}} \Bigr)^2 \, ,\nonumber\\
c_z &=& \frac{C}{A}\Bigl( \frac{L_{\bar{x}}}{L_{\bar{z}}} \Bigr)^2 \, ,
\label{coefs}\\
E &=& E_{\text{r}} + i E_{\text{i}} =
\frac{\bar E}{\alpha_t} \, .
\end{eqnarray}

\section{Numerical algorithms}

Integrating the evolution equation (\ref{eveq}) from some value of $t$ to a
later instant $t + \Delta t$, $\Delta t > 0$, we obtain
\begin{equation}\label{evres}
u(t + \Delta t) = u(t) + \left. \int_t^{t + \Delta t} F[u]\right|_{\tau} \,
d \tau \, .
\end{equation}
Assuming that $\Delta t$ is small, simple numerical algorithms and their error
estimates can be obtained from (\ref{evres}) by using the parabolic
interpolation formula for $F[u]|_{\tau}$:
\begin{equation}\label{interp}
F[u]\rvert_{\tau} = F_0 + \beta (\tau - t) +
\gamma (\tau - t)[\tau - (t + \Delta t)] \, ,
\end{equation}
where the quantities $\beta$ and $\gamma$ can be expressed in terms of
\begin{equation}\label{f0}
F_0 = F[u]|_t \, , \quad F_1 = F[u]|_{t + \Delta t} \, , \quad
F_{1/2} = F[u]|_{t + \Delta t/2} \, ,
\end{equation}
i.e.
\begin{equation}\label{ag}
\beta = \frac{F_1 - F_0}{\Delta t} \, , \qquad
\gamma = \frac{2}{(\Delta t)^2} [F_0 - 2 F_{1/2} + F_1] \, .
\end{equation}
$\beta$ and $\gamma$ are nearly independent of $\Delta t$
(close to their limits as $\Delta t \to 0$).

Inserting (\ref{interp}) into (\ref{evres}) we can find two algorithms,\\
an Euler algorithm (first order in $\Delta t$):
\begin{equation}\label{euler}
u(t + \Delta t) = u(t) + \Delta t F_0 + O[(\Delta t)^2] \, ,
\end{equation}
a Trapez algorithm (second order in $\Delta t$):
\begin{equation}\label{implic}
u(t + \Delta t) = u(t) + \Delta t \, {\textstyle \frac{1}{2}} (F_0 + F_1) +
O[(\Delta t)^3] \, .
\end{equation}
Note that (\ref{implic}) is an implicit algorithm: $u(t + \Delta t)$ is
defined in terms of $u(t)$ and $u(t + \Delta t)$, i.e. equation
(\ref{implic}) must be solved for $u(t + \Delta t)$. Usually, this is done
by an iterative procedure, where in the lowest approximation $F_1$ on the
RHS of (\ref{implic}) is replaced by $F_0$. This defines the first
approximation to $u(t + \Delta t)$ to be used on the RHS of (\ref{implic})
to define second approximation to $u(t + \Delta t)$, etc.

Another simple (explicit) algorithm can be obtained from (\ref{implic}) if
we put $\tau = t + \Delta t/2$ in (\ref{interp}):
\[
F_{1/2} = {\textstyle \frac{1}{2}} (F_0 + F_1) - \gamma (\Delta t)^2/4 \, ,
\]
and use ${\textstyle \frac{1}{2}} (F_0 + F_1)$ calculated from this equation in
(\ref{implic}):
\begin{equation}\label{lfrg0}
u(t + \Delta t) = u(t) + \Delta t F_{1/2} + O[(\Delta t)^3] \, .
\end{equation}
This formula defines a two-point algorithm: $u(t + \Delta t)$ is given in
terms of $u(t)$ and $u(t + \Delta t/2)$, i.e. at the centre of the interval
$[t, \, t + \Delta t]$.
Both $u(t)$ and $u(t + \Delta t/2)$ must be known, and hence the actual
evolution interval is $\Delta t/2$ rather than $\Delta t$. Replacing in
(\ref{lfrg0}) $u(t) \to u(t - \Delta t)$, the central point will now be at
$t$, and the integration interval will be $2 \Delta t$:
\begin{equation}\label{lfrg}
u(t + \Delta t) = u(t - \Delta t) + 2 \Delta t F_0 + O[(\Delta t)^3] \, .
\end{equation}
Here $u(t + \Delta t)$ is defined in terms of $u(t - \Delta t)$ and $u(t)$.
Using the known value $u(t)$, and $u(t + \Delta t)$ calculated from
(\ref{lfrg}) we can determine $u(t + 2 \Delta t)$, etc. (the leapfrog
procedure).
The only problem is to start this procedure, which requires two values of $u$:
$u(t_0)$ and $u(t_0 + \Delta t)$, where $t_0$ is an initial value of $t$.
But if $u(t_0)$ is prescribed, the evolution equation (\ref{eveq}) (of first
order in $t$) defines $u(t)$ for any $t > t_0$. In particular, $u(t_0 +
\Delta t)$ is defined. It can be calculated up to $O[(\Delta t)^2]$ by
using the Euler algorithm (\ref{euler}). Using this approximation to
determine $F[u]$ required for (\ref{lfrg}), the error in $F[u]$ will also
be $O[(\Delta t)^2]$. After multiplication by $\Delta t$ it will produce
an error comparable with that in (\ref{lfrg}), $O[(\Delta t)^3]$.

In the pseudospectral method described  in \cite{fornwh}, the Discrete Fourier Transform
(DFT) with respect to each space variable is used to calculate the derivatives in
(\ref{rhs}). Thus the interval $[0,2 \pi]$ for $x$ will be divided  into  $N_x$
subintervals of length $\Delta x = 2 \pi/N_x$, and  similarly
for $y$ and $z$ ($N_y$ subintervals of length $\Delta y = 2 \pi/N_y$
and $N_z$ subintervals of length $\Delta z = 2 \pi/N_z$; the numbers $N_x$,
$N_y$ and $N_z$ can be either even or odd, $N_x=2 M_x$ or $N_x=2 M_x + 1$, etc.). The
function $u$ defined on the discrete mesh $(x_j, y_m, z_n), \,
x_j = j \Delta x, \, y_m = m \Delta y, \, z_n = n \Delta z$, can  be 
transformed to  discrete  Fourier  space  for $x$, $y$, and $z$ 
variables. Thus for each $y_m$ and $z_n$  we  define  the  Discrete  Fourier 
Transform in $x$:
\begin{equation}
v(k_x) = \frac{1}{\sqrt{N_x}} \sum_{j=0}^{N_x - 1} u(x_j) \exp(
- i k_x x_j) \, .\label{dft}\\
\end{equation}           
The inverse transform is given by
\begin{equation}
u(x_j) = \frac{1}{\sqrt{N_x}} \left. \sum_{k_x=-M_x}^{M_x} v(k_x)
\exp(i k_x x) \right\rvert_{x=x_j} \, ,\label{invdft}
\end{equation}
both if $N_x=2 M_x$ or $N_x=2 M_x + 1$. In the first case, the summation index in
(\ref{invdft}) actually ends up with $k_x = M_x - 1$. However, as in that case $v(k_x)
\exp(i k_x x_j)$ is periodic as function of $k_x$ with period $2 M_x$, Eq.~(\ref{invdft})
will be correct if only one half of the contributions at $k_x = \pm M_x$ are included in
the sum over $k_x$. Replacing $x \to y,z$ and $j \to m,n$  everywhere in (\ref{dft}) and
(\ref{invdft}) we obtain the formulas for the  DFT in $y$ (for each $x_j$ and $z_n$)
and in $z$ (for each $x_j$ and $y_m$). The  essence  of the  pseudospectral 
approach is  to  calculate the  partial derivatives at the mesh  points  by 
differentiating  the interpolation formula (\ref{invdft})  with respect 
to $x$ (or its analogues with respect to $y$ or $z$). Thus, for example
\begin{equation}
u_{xx}(x_j) = \frac{1}{\sqrt{N_x}}
\sum_{k_x} ( i k_x)^2 v(k_x) \exp( i k_x x_j)
\, ,\label{uxx}
\end{equation}
etc. In our numerical programs, the sums of the type (\ref{dft}) or
(\ref{uxx}) were determined by using either the Fast Fourier Transform (FFT)
subroutine in complex domain, given in \cite{numrec} (in the early version
of the program from 1996),
or more efficient ``Multiple 1D FFT subroutines for complex data'', taken
from the NAG Fortran Library (in the later versions from 1999 and 2004).

The Discrete Fourier Transform as described above is an efficient tool for numerical
differentiation. The only problem is that Eq.~(\ref{invdft}) implies periodicity of
the function $u(x)$ which may not be the case, neither exactly nor approximately. If
that happens, one can follow the suggestion of P. J. Roach \cite{roach} to split the
function $u(x)$ into the periodic component given by (\ref{invdft}) and a non-periodic
one in the form of a polynomial. Another approach could be to replace the DFT by
expansion of $u(x)$ in a non-periodic orthogonal basis, e.g. that of orthogonal
polynomials. This type of approach using the Chebyshev polynomials
has recently been discused in detail and tested by A. Deloff \cite{deloff}.

\section{Numerical stability}

To examine the numerical stability of the algorithms derived in the previous
section, we linearize $F[u]$, Eq.~(\ref{rhs}), by  replacing
\begin{equation}\label{linear}
|u|^2 u \to |u_0|^2 u, \quad u_0 = \text{const} \, .
\end{equation}
As Eqs.~(\ref{euler}), (\ref{implic}) and
(\ref{lfrg}) (without error terms) have constant coefficients, their solutions can be
looked for in the form of exponential functions of $x$, $y$, $z$ and $t$:
\begin{equation}\label{expsol}
u = \kappa^{t/\Delta t} \exp[ i (k_x x + k_y y + k_z z)] \, .
\end{equation}
Numerical stability of the algorithm in question means that the solution
(\ref{expsol}) cannot grow in time, i.e., $|\kappa| \le 1$.

Inserting (\ref{expsol}) into (\ref{rhs}) we obtain
\begin{eqnarray}
F[u] & = & i \bigl[w + i E_{\text{i}}\bigr] u \, ,\label{Fu}\\
   w & = & - \text{sgn}(A) (k_x^2 + c_y k_y^2 + c_z k_z^2)\nonumber\\
&&+ \text{sgn}(D)|u_0|^2 + E_{\text{r}} \, ,\label{w}
\end{eqnarray}
and furthermore
\begin{equation}\label{upmdt}
u(t + \Delta t) = u \kappa \, , \qquad u(t - \Delta t) = u/\kappa \, .
\end{equation}
Eq.~(\ref{w}) leads to
\begin{equation}\label{max}
|w|_{\text{max}} \leq M_x^2 + |c_y| M_y^2 + |c_z| M_z^2 + |u_0|^2 + |E_{\text{r}}| \, ,
\end{equation}
where $M_x = \text{max} |k_x|$, etc., see Eq.~(\ref{invdft}).
This estimate is an accurare expression for $|w|_{\text{max}}$ if all
terms in (\ref{w}) have the same sign. Otherwise certain terms in (\ref{max})
should be discarded. Nevertheless, in practice the overestimation given by
the RHS of (\ref{max}) in that case is not large, and if this formula is used
in the expressions for maximal $\Delta t$ for stability given in what follows,
the only effect will be the introduction of a small safety margin. To obtain
a precise expression for $|w|_{\text{max}}$, all terms in (\ref{w}) should
be divided into two groups, of positive and negative. The  contribution
to the RHS of (\ref{max}) of these two groups should be compared, and the
group of terms with smaller contribution should be discarded.

\subsection{Euler algorithm}

Using (\ref{Fu}) and (\ref{upmdt}) in (\ref{euler}) we obtain
\begin{equation}\label{keuler}
\kappa = 1 + i \Delta t [w + i E_{\text{i}}] \, ,
\end{equation}
i.e.
\begin{eqnarray}
|\kappa|^2 &=& 1 - 2 \Delta t E_{\text{i}} + (\Delta t)^2
[E_{\text{i}}^2 + w^2]\nonumber\\
&=& 1 - \Delta t [2 E_{\text{i}} - \Delta t
(E_{\text{i}}^2 + w^2)] \, .\label{mkeuler}
\end{eqnarray}
Thus if $E_{\text{i}} \leq 0$, we obtain $|\kappa|^2 > 1$, i.e. the Euler
algorithm is numerically unstable.

Numerical stability is only possible for $E_{\text{i}} > 0$ if
\[ 
0 < \Delta t [2 E_{\text{i}} - \Delta t (E_{\text{i}}^2 + w^2)] < 1 \, .
\]
As in practice $|w|_{\text{max}} \gg |E_{\text{i}}|$, the stability condition
for the Euler algorithm takes the form
\begin{equation}\label{eulerst}
E_{\text{i}} > 0 \quad \text{and} \quad \Delta t <
\frac{2 E_{\text{i}}}{E_{\text{i}}^2 +
(|w|_{\text{max}})^2} \, ,
\end{equation}
where $|w|_{\text{max}}$ is given by (\ref{max}) (with a possible modification
as described above).

\subsection{Implicit algorithm}

Using (\ref{Fu}) and (\ref{upmdt}) in (\ref{implic}) we obtain
\begin{equation}\label{kimplic}
\kappa = \frac{1 - {\textstyle \frac{1}{2}} \Delta t (E_{\text{i}} - i w)}%
{1 + {\textstyle \frac{1}{2}} \Delta t (E_{\text{i}} - i w)} \, ,
\end{equation}
i.e.
\begin{equation}\label{mkimplic}
|\kappa|^2 = 1 - p \, , \qquad
p = \frac{2 \Delta t E_{\text{i}}}%
{(1 + {\textstyle \frac{1}{2}} \Delta t E_{\text{i}})^2 +
({\textstyle \frac{1}{2}} \Delta t w)^2}
\, .
\end{equation}
Thus if $E_{\text{i}} < 0$, we obtain $|\kappa|^2 > 1$, i.e. the implicit
algorithm is
numerically unstable, and for $E_{\text{i}} = 0$ this algorithm is marginally
stable
($|\kappa|^2 = 1$). And finally, for $E_{\text{i}} > 0$, $p$ is positive and
should not
be greater than one (again due to expected $|w|_{\text{max}} \gg
|E_{\text{i}}|$), which means numerical stability for any value of $\Delta t$.

\subsection{Leapfrog algorithm}

Using (\ref{Fu}) and (\ref{upmdt}) in (\ref{lfrg}) we obtain a quadratic in $\kappa$
\begin{equation}\label{quadr}
\kappa^2 + 2 \Delta t (E_{\text{i}} - {\text{i}} w) \kappa - 1 = 0 \, .
\end{equation}
Solving Eq. (\ref{quadr}) we obtain
\begin{equation}\label{klfrg}
\kappa = - \Delta t (E_{\text{i}} - i w) \pm
\sqrt{1 + (\Delta t)^2 (E_{\text{i}} - i w)^2} \, .
\end{equation}
For $E_{\text{i}} \neq 0$, the general expressions for the real and
imaginary part of
$\kappa$ are a bit complicated, but in the limit $\Delta t \to 0$ we easily
find
\begin{equation}\label{ksmdt}
\kappa \simeq \pm 1 - \Delta t (E_{\text{i}} + i w)
\, , \quad \text{i.e.} \quad |\kappa|^2 \simeq 1 \pm 2 \Delta t E_{\text{i}} \, .
\end{equation}
This can always be greater than one, which means instability.

For $E_{\text{i}} = 0$, we obtain
\begin{equation}\label{k0}
\kappa = - i a \pm \sqrt{1 - a^2} \, , \qquad a = \Delta t \, w \, .
\end{equation}
Hence if $|a| \le 1$ we obtain $|\kappa|^2 \equiv 1$, which means marginal
stability, whereas for $|a| > 1$ we get
\[
|\kappa|_{\text{max}} = |a| + \sqrt{a^2 - 1} > |a| > 1 \, ,
\]
which means instability. Hence the numerical stability condition is
$|a| \le 1$. This condition leads to the following formula for
maximal timestep $\Delta t$ for stability:
\begin{equation}\label{maxdtl}
\Delta t = \frac{c}{|w|_{\text{max}}} \, .
\end{equation}
where $0 < c < 1$, and $|w|_{\text{max}}$ is given by (\ref{max}) (again
with the possible modification).

\begin{acknowledgments}
This research was supported by the Committee for Scientific Research (KBN),
Grant No KBN 2P03B09722.
\end{acknowledgments}

\end{document}